\documentclass[12pt,preprint]{aastex}

\slugcomment{Submitted to Astrophysical Journal Letters}

\shorttitle{Jovian Meridional Transport}

\shortauthors{Liang et al.}

\begin{document}

\title{Meridional Transport in the Stratosphere of Jupiter}

\author{Mao-Chang Liang$^{1*}$, Run-Lie Shia$^{1}$, Anthony Y.-T. Lee$^{1}$, Mark Allen$^{1,2}$, A. James Friedson$^{2}$, and Yuk L. Yung$^{1}$}

\affil{$^{1}$Division of Geological and Planetary Sciences,
California Institute of Technology, Pasadena, CA 91125}

\affil{$^{2}$Jet Propulsion Laboratory, California Institute of
Technology, Pasadena, CA 91109}

\email{$^*$To whom all correspondence should be addressed. E-mail:
mcl@gps.caltech.edu}

\begin{abstract}
The Cassini measurements of C$_2$H$_2$ and C$_2$H$_6$ at $\sim$5
mbar provide a constraint on meridional transport in the
stratosphere of Jupiter. We performed a two-dimensional
photochemical calculation coupled with mass transport due to
vertical and meridional mixing. The modeled profile of C$_2$H$_2$ at
latitudes less than 70$^\circ$ follows the latitude dependence of
the solar insolation, while that of C$_2$H$_6$ shows little latitude
dependence, consistent with the measurements. In general, our model
study suggests that the meridional transport timescale above 5-10
mbar altitude level is $\gtrsim$1000 years and the time could be as
short as 10 years below 10 mbar level, in order to fit the Cassini
measurements. The derived meridional transport timescale above the 5
mbar level is a hundred times longer than that obtained from the
spreading of gas-phase molecules deposited after the impact of
Shoemaker-Levy 9 comet. There is no explanation at this time for
this discrepancy.
\end{abstract}


\section{Introduction}
Meridional advection and mixing processes in the atmosphere of
Jupiter are poorly known. Based on the Voyager infrared spectrometer
data, several efforts to derive the atmospheric circulation have
been published \citep[e.g.,][]{Gierasch et al.1986,Conrath et
al.1990,West et al.1992}. The first direct and quantitative
derivation of meridional transport processes is based on the
introduction of aerosol debris into the atmosphere of Jupiter by
comet Shoemaker-Levy 9 (SL9). \citet{Friedson et al.1999} conclude
that the advection by the residual circulation calculated by
\citet{West et al.1992} is insufficient to explain the temporal
distribution of cometary debris. Meridional eddy mixing coefficients
on the order of 1-10$\times$10$^{10}$ cm$^2$ s$^{-1}$ are inferred
in the regions between $\sim$10 and 100 mbar. Later, based on the
time evolution profiles of CO, CO$_2$, CS, HCN, and H$_2$O gas-phase
molecules deposited after the SL9 impact, values of meridional eddy mixing coefficients as high as
2-5$\times$10$^{11}$ cm$^2$ s$^{-1}$ are derived for pressures
between $\sim$0.1-0.5 mbar \citep{Lellouch et al.2002,Moreno et
al.2003,Griffith et al.2004}.

The Cassini measurements of stratospheric C$_2$H$_2$ and C$_2$H$_6$
\citep{Kunde et al.2004} provide good tracers for characterizing
mass transport in the upper atmosphere of Jupiter. However, the
weighting function of these observations is such that they are most
sensitive to the altitude level near 5 mbar. Kunde et al. show that,
at latitudes equatorward of $\sim$70$^\circ$, the relative magnitude
of the abundance (or emission line intensity) of C$_2$H$_2$ follows
the latitudinal variation in solar insolation, while the abundance
(or emission line intensity) of C$_2$H$_6$ is constant with
latitude. Consequently, Kunde et al. conclude that the stratospheric
meridional transport timescale at latitudes $<$70$^\circ$ derived
from these Cassini data falls between the lifetimes of C$_2$H$_2$
and C$_2$H$_6$. Due to the complexity in the auroral regions
(contamination of line emissions from higher atmosphere due to
temperature enhancement), we focus on the regions with latitude less
than $\sim$70$^\circ$ in this paper.

\section{Two-Dimensional Transport Model}
Because of the rapid rotation of Jupiter and its strong
stratospheric zonal wind, the zonal variations in abundances should
be minimized\footnote{The rotation period and atmosphere-radiative
timescale is $\sim$10 hours and $\sim$1000 days (at $\sim$10 mbar)
\citep{Flasar1989}, respectively.}. Therefore, a two-dimensional
(2-D) model including net meridional and vertical transport should
be an adequate first-order simulation of these Cassini observations
of the trace hydrocarbon species. In the 2-D mode, the Caltech/JPL
coupled chemistry/transport code, solves the mass continuity
equation:
\begin{eqnarray} \frac{{\partial
n_i(y, z, t)}}{{\partial t}} + \nabla\cdot\vec{\varphi_i}(y, z, t) &
= & P_i(y, z, t) - L_i(y, z, t),
\end{eqnarray}
where $n_i$ is the number density for the species $i$, $\varphi_i$
the transport flux, $P_i$ the chemical production rate, and $L_i$
the chemical loss rate, all evaluated at time $t$, latitudinal
distance $y$ and altitude $z$. Reported herein are results for
diurnally averaged steady-state solutions, i.e., $<$$\partial n_i /
\partial t$$>$ $\rightarrow 0$.

For one-dimensional (1-D) problems ($y$ dependence in eqn.(1)
vanishes), the Caltech/JPL code \citep[see, e.g.,][]{Gladstone et
al.1996} integrates the continuity equation including chemistry and
vertical diffusion for each species using a matrix inversion method
that allows large time steps. To take advantage of the computational
efficiency of the 1-D solver, a ``quasi 2-D" mode, which is a series
of 1-D models at different latitudes coupled by meridional
transport, has been developed. This quasi 2-D simulation has been
tested against a case which has a known solution \citep{Shia et
al.1990}. Since the meridional transport in the stratosphere is not
well understood, the quasi 2-D model uses a simple parameterization
for mixing between the neighboring 1-D columns to simulate the
meridional transport, i.e., $K_{yy}$$\partial n_i$/$\partial y$ is
added to the results of the 1-D computations, where $K_{yy}$ is the
meridional mixing coefficient. Our work provides an order of
magnitude estimate of the meridional transport in the stratosphere.

The vertical mixing coefficients ($K_{zz}$) and temperature profile
are taken from \citet{Gladstone et al.1996} and are assumed to be
independent of latitude in the reference model. Sensitivities of the
results due to the variations of temperature and $K_{zz}$ profiles
are also presented.

The meridional mixing $K_{yy}$ will be determined by fitting the
Cassini measurements \citep{Kunde et al.2004}. As a first order
approximation, the $K_{yy}$ profile will be assumed also to be
latitude-independent.

The model atmosphere is gridded latitudinally and vertically in 10
and 131 layers, respectively. The vertical grid size is chosen to
insure that there are $>$3 grid points in one pressure scale height
(to achieve good numerical accuracy). The latitude grid points are
at 10, 30, 50, 70, and 85$^\circ$ in each hemisphere.

The photochemical reactions ($P_i$ and $L_i$) are taken from
\citet{Moses et al.2000}. At all latitudes, the mixing ratio of
CH$_4$ in the deep atmosphere is prescribed to be
2.2$\times$10$^{-3}$, and the atomic hydrogen influx from the top
atmosphere is fixed at 4$\times$10$^9$ cm$^{-2}$ s$^{-1}$
\citep[see, e.g.,][]{Gladstone et al.1996}. In order to prevent the
solution from oscillating seasonally, the inclination angle of
Jupiter is prescribed to be zero ($\sim$3$^\circ$ actually).

\section{Simulation Results}
\begin{figure*}
 \epsscale{1} \plotone{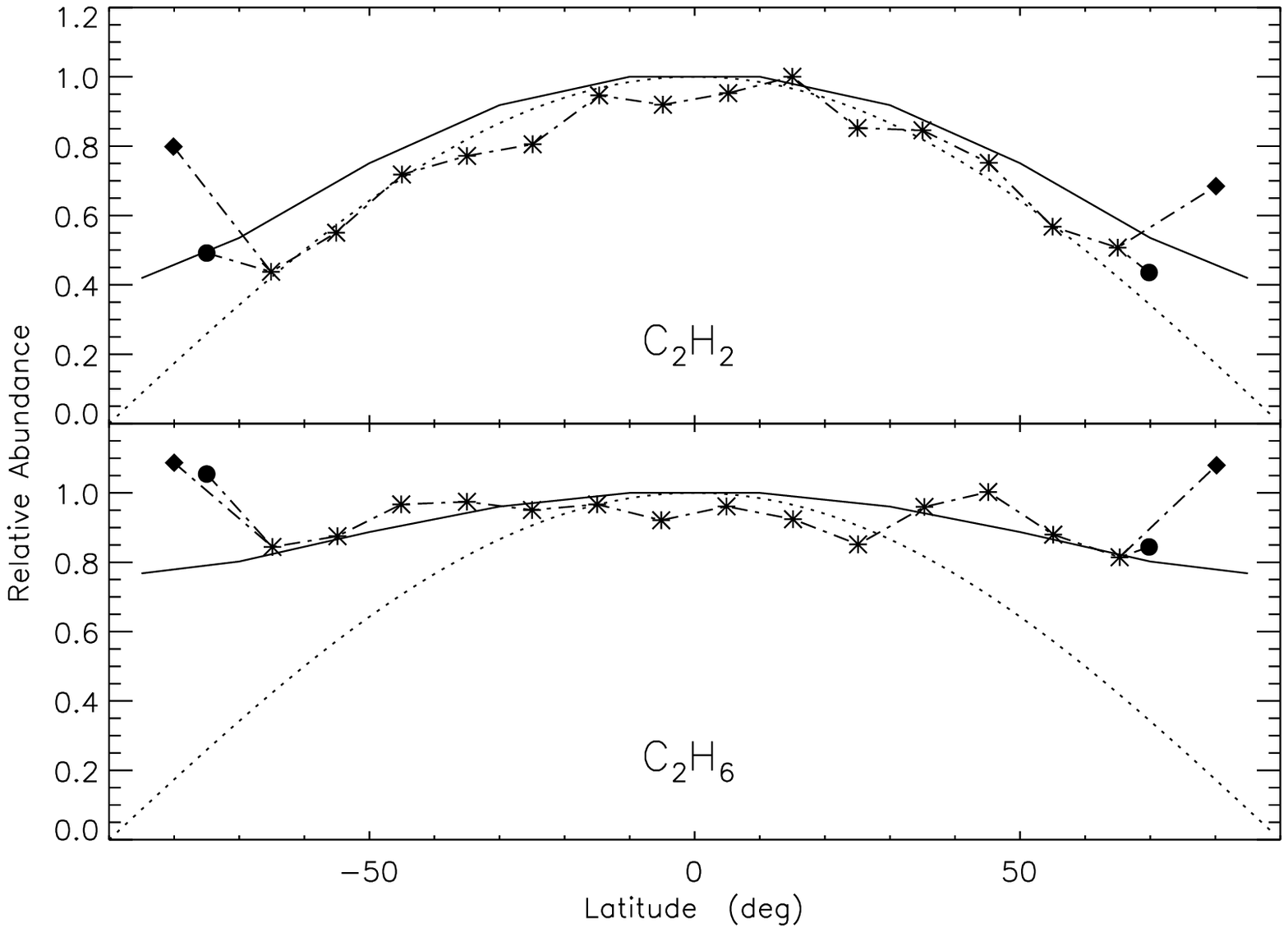}
\caption[Profiles of C2H2 and C2H6 at 5 mbar]{Relative abundances of
C$_2$H$_2$ (upper) and C$_2$H$_6$ (lower) at 5 mbar as a function of
latitude. Asterisks are the Cassini measurements \citep{Kunde et
al.2004}. Diamonds and circles are, respectively, high-latitude
Cassini measurements that do include (diamonds) and do not include
(circles) auroral longitudes. The calculated abundances of
C$_2$H$_2$ and C$_2$H$_6$ are normalized to those at the equator.
Solid line represents model results with the reference $K_{yy}$:
constant 2$\times$10$^{10}$ cm$^{2}$ s$^{-1}$ below the 5 mbar
altitude level and 2$\times$10$^{9}$ cm$^{2}$ s$^{-1}$ above (model
B). Dotted line represents the cosine function of latitude.
\label{c2profiles}}
\end{figure*}

The Cassini measurements of the latitudinal distribution of
C$_2$H$_2$ and C$_2$H$_6$ are reproduced in Fig.~\ref{c2profiles};
the contribution function for these is near 5 mbar \citep{Kunde et
al.2004}. Since the absolute abundances of C$_2$H$_2$ and C$_2$H$_6$
have not been determined from these measurements, the results shown
are arbitrarily scaled following the approach of Kunde et al.
Because the C$_2$H$_2$ abundances follow the latitudinal
distribution of insolation (insolation is proportional to cosine of
latitude), the meridional mixing timescale is expected to be longer
than the C$_2$H$_2$ photochemical lifetime. On the other hand, since
the C$_2$H$_6$ abundance appears to be constant with latitude, the
meridional mixing timescale is expected to be shorter than the
C$_2$H$_6$ photochemical lifetime.

\begin{figure*}
 \epsscale{1} \plotone{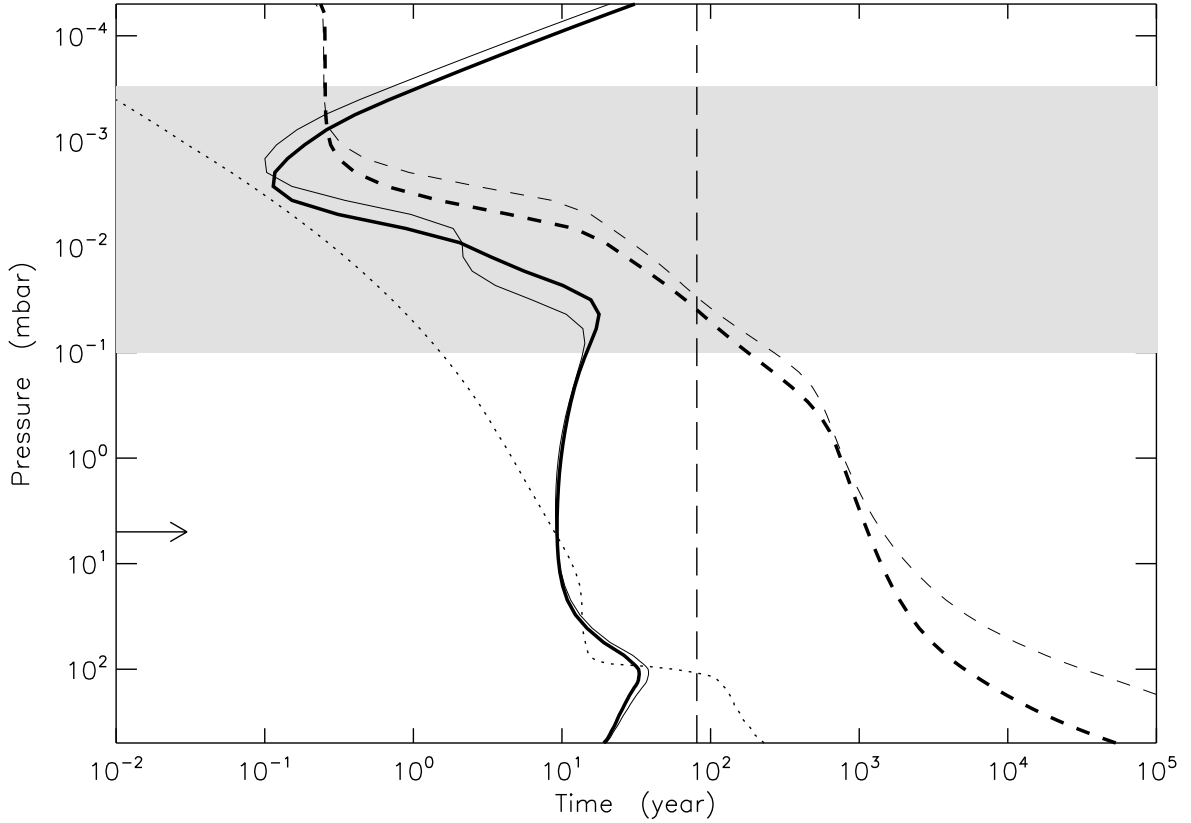}
\caption[Profiles of some time constants]{Timescales for the
chemical loss of C$_2$H$_2$ (solid lines) and C$_2$H$_6$ (dashed
lines) and for vertical transport (dotted line). The vertical
transport timescale is defined by $H^2$/$K_{zz}$, where $K_{zz}$ and
$H$ are the vertical diffusion coefficients of CH$_4$ and
atmospheric scale height, respectively; values of time constants are
derived from our reference model. Thick and thin lines represent
values at latitudes 10$^\circ$ and 70$^\circ$, respectively. The
horizontal arrow indicates 5 mbar level where peak in the
contribution function for the Cassini measurements of C$_2$H$_2$ and
C$_2$H$_6$ lies. The vertical long-dashed line is a meridional
mixing time equal to $R_J^2$/$K_{yy}$, where $R_J$ is the radius of
Jupiter and $K_{yy}$ = 2$\times$10$^{10}$ cm$^2$ s$^{-1}$. The
shaded area shows the photochemical production region for the
hydrocarbons. \label{time}}
\end{figure*}

A more quantitative estimate for $K_{yy}$ can be derived from the
2-D model. A validation of the 2-D model is from a 1-D model for
Jovian hydrocarbon chemistry \citep{Gladstone et al.1996,Moses et
al.2005}, which reproduces extensive observations of hydrocarbon
species as well as He 584 \AA\ and H Lyman-$\alpha$ airglow
emissions at low latitudes (observations are summarized in the
Tables 1 and 5 of Gladstone et al. 1996 and Fig. 14 of Moses et al.
2005). The chemical loss timescales for C$_2$H$_2$ and C$_2$H$_6$
drawn from our current work are shown in Fig.~\ref{time}, along with
vertical mixing timescales (also from our current work). At 5 mbar,
the chemical loss timescales for C$_2$H$_2$ and C$_2$H$_6$ are about
10 and 2000 years, respectively. Therefore, to reproduce the Cassini
distributions of C$_2$H$_2$ and C$_2$H$_6$, the meridional mixing
time must fall within 10 and 2000 years at 5 mbar.

Many simulations were performed with the quasi-2D model to explore
the sensitivities of the abundances C$_2$H$_2$ and C$_2$H$_6$ at 5
mbar to the choice of $K_{yy}$. The results were parameterized in
terms of the ratio of the abundance at 70$^\circ$ latitude to the
abundance at 10$^\circ$. In general, the altitude variation of
$K_{yy}$ leading to a latitudinal gradient for C$_2$H$_2$ consistent
with the Cassini observations also led to a sharp reduction in the
C$_2$H$_6$ abundance from equator to near-pole. Alternatively, for
many cases tested, if the C$_2$H$_6$ equator to near-pole variation
was small, the same was true for the C$_2$H$_2$ equator to near-pole
variation. Shown in Table 1 are the model results that most
adequately reproduce the Cassini observations. We found that a
`transition' level somewhere around 5 and 10 mbar must be present,
in order to match the Cassini measurements. Above the transition
altitude, $K_{yy}$ $\lesssim$10$^9$ cm$^2$ s$^{-1}$. Below the
transition altitude, $K_{yy}$ $>$
2$\times$10$^{10}$ cm$^2$ s$^{-1}$, consistent
with the analysis of the temporal spreading of the SL9 debris
\citep{Friedson et al.1999}. It is interesting to note that the
results of Friedson et al. predict a reversal in the direction of
the meridional component of the annual-mean residual velocity across
the 5 mbar level in the southern hemisphere.  This reversal
represents a change with altitude in the relative contributions to
the total annual-mean meridional heat flux from the component
associated with the Eulerian-mean meridional velocity and the
component associated with eddies. It is therefore possible that a
change in the efficiency of meridional transport accompanies the
reversal.

These model results can be understood in the context of the basic
photochemistry controlling the production and loss of C$_2$H$_2$ and
C$_2$H$_6$ as outlined in \citet{Gladstone et al.1996}. In the
atmosphere of Jupiter, most of hydrocarbon compounds are synthesized
in the regions above $\sim$0.1 mbar (shown in the shaded area in
Fig.~\ref{time}). Below 0.1 mbar, the production of C$_2$H$_2$
through CH$_4$ photolysis mediated reactions is insufficient,
because the process requires UV photons with wavelengths $<$130 nm,
which has been self-shielded by CH$_4$. In this region, the
photolysis of C$_2$H$_6$ ($<$160 nm), which decreases strongly
toward lower regions of the atmosphere, dominates the production of
C$_2$H$_2$. The major chemical loss of C$_2$H$_2$ and C$_2$H$_6$ is
through hydrogenation and photolysis, respectively. As a consequence
of this chemistry, C$_2$H$_2$ is close to being in photochemical
steady-state in the regions below $\sim$5 mbar altitude level and
its vertical gradient is large in this altitude range. Above 5 mbar
level, transport is important for C$_2$H$_2$ (Fig.~\ref{time}). On the other hand, the
abundance of C$_2$H$_6$ is controlled by transport and its vertical
gradient is very small (see, e.g., Fig.~14 in Gladstone et al.
1996). Therefore, if meridional mixing is sufficiently rapid below
the transition altitude to uniformly mix C$_2$H$_6$ with latitude,
the tendency for uniform latitudinal mixing occurs also at 5 mbar.
While above the transition altitude, uniform latitudinal mixing of
C$_2$H$_6$ also results in uniform latitudinal mixing of C$_2$H$_2$
at 5 mbar.

Fig.~\ref{2D} shows the 2-D distribution of C$_2$H$_2$ and
C$_2$H$_6$ calculated with the reference model, in which $K_{yy}$ =
2$\times$10$^{10}$ below 5 mbar level and 2$\times$10$^{9}$ cm$^2$
s$^{-1}$ above. Additional observations at different levels in the
atmosphere can help constrain the 2-D dynamical properties of the
Jovian atmosphere.

vertical profiles of C$_2$H$_2$ and C$_2$H$_6$ at two latitudes. For
comparison, 1-D results at low latitude are also overplotted.
Additional observations at different levels in the atmosphere can
help constrain the 2-D dynamical properties of the Jovian
atmosphere.

In the above calculations whose results are summarized in Table
\ref{sensitivity}, assumptions were made with respect to temperature
and vertical eddy diffusion coefficients independent of latitudes
and in the selection of chemistry reaction parameters. Several
additional calculations were performed to assess the sensitivity of
the derived values for $K_{yy}$ with respect to these assumptions.
In one sensitivity test, the temperature profile was progressively
increased from 10$^\circ$ latitude so that, by 85$^\circ$, the
temperature profile was 10\% larger. Table \ref{sensitivity}, model
G, shows that, with the adjusted temperature distribution, the
$K_{yy}$ that best simulated the latitudinal variations in
C$_2$H$_2$ and C$_2$H$_6$ in the Cassini observations is the same as
derived above. This is consistent with the conclusions of
\citet{Moses and Greathouse2005} who found little sensitivity to
temperature in their calculations. In a similar fashion, $K_{zz}$
was modified linearly so that the value at 85$^\circ$ latitude was
10 times lower than the value at 10$^\circ$ latitude; enhanced
$K_{zz}$ at high latitudes cannot reproduce the measurements. The
$K_{yy}$ values that best simulated the Cassini observations (Table
\ref{sensitivity}, model H) were again those derived earlier in this
paper. Finally, the same result for $K_{yy}$ was found when the
model chemistry was updated to be consistent with the reaction
coefficients in \citet{Moses et al.2005} (Table \ref{sensitivity},
model F). Therefore, the conclusions in this paper for the magnitude
of meridional mixing as a function of altitude are robust with
regard to reasonable selection of atmospheric temperature, vertical
mixing, and chemistry.

\section{Conclusion}
Our model simulation results have two implications. First, the
meridional transport time as short as 10 years ($K_{yy}$ $\approx$
10$^{11}$ cm$^2$ s$^{-1}$) exists only in the altitude range below
the 10 mbar level. Second, above a transition level somewhere
between 5 and 10 mbar, the meridional transport time is not shorter
than $\sim$1000 years ($K_{yy}$ $\lesssim$ 10$^9$ cm$^2$ s$^{-1}$).
While these inferred $K_{yy}$ values for the atmosphere at and below
the 5-10 mbar level are consistent with the conclusions of
\citet{Friedson et al.1999} derived from an analysis of the SL9
debris evolution with time, the $K_{yy}$ value above the transition
level is much smaller than that ($\sim$10$^{11}$ cm$^2$ s$^{-1}$)
derived from analysis of the time evolution of the distributions of
gas-phase trace species deposited after the SL9 impact
\citep{Lellouch et al.2002,Moreno et al.2003,Griffith et al.2004}.
There is no explanation at this time for this discrepancy.

\begin{figure*}
 \epsscale{1} \plotone{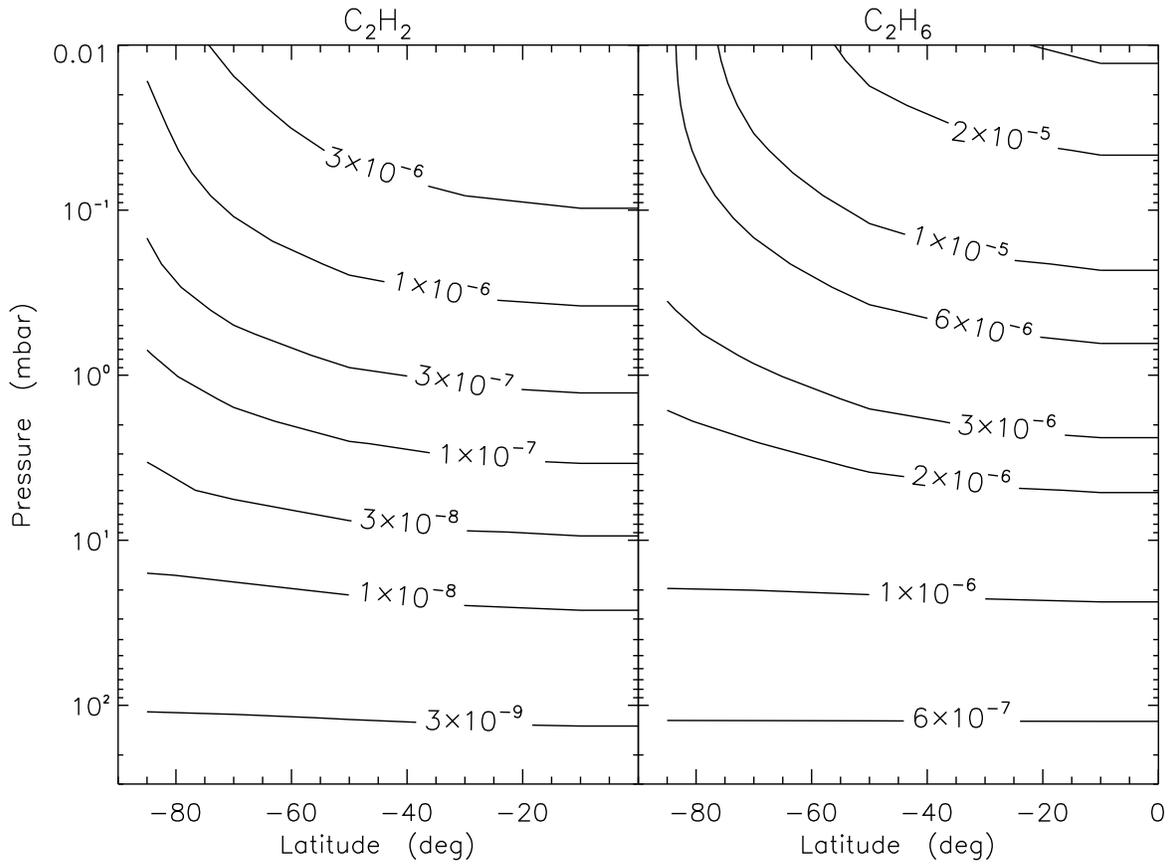}
\caption[2D plots of C2H2 and C2H6]{2-D volume mixing ratio profiles
of C$_2$H$_2$ (left) and C$_2$H$_6$ (right) calculated with the
reference $K_{yy}$ (model B). \label{2D}}

\end{figure*}

It has been shown that CH$_4$ and C$_2$H$_6$ contribute to the
heating and cooling of the stratosphere of Jupiter, respectively
\citep{Yelle et al.2001}. To a first order approximation, the
cooling at/below 5 mbar level would be constant with latitude, being
determined by two factors, the temperature and the abundance of
C$_2$H$_6$. These two are nearly constant between the equator and
mid-latitudes \citep[also Fig.~\ref{2D}]{Flasar et al.2004,Kunde et
al.2004}. The heating function, however, is sensitive to the
magnitude of the solar insolation and, consequently, it is a good
approximation to assume that this function has the same latitude
dependence as the solar insolation. Therefore, the global
circulation could be driven by the differential heating between
latitudes. This circulation driven by heating through absorption of
radiation by gas-phase molecules will provide a first order estimate
of the importance of aerosol heating in the stratosphere.

\acknowledgements This research was supported in part by NASA grant
NAG5-6263 to the California Institute of Technology. Special thank
to Julie Moses for her insightful comments.

\clearpage

\clearpage

\clearpage

\clearpage

\begin{deluxetable}{llllrrrrr}
\tabletypesize{\scriptsize} \tablecolumns{7} \tablecaption{Summary
of model results \label{sensitivity}} \tablewidth{0pt} \tablehead{
\multicolumn{1}{c}{} &
\multicolumn{1}{c}{Chemistry\tablenotemark{a}} &
\multicolumn{1}{c}{Temperature\tablenotemark{a}} &
\multicolumn{1}{c}{$K_{zz}$\tablenotemark{a}}
&\multicolumn{1}{c}{Transition\tablenotemark{b}} &
\multicolumn{1}{c}{$K_{yy}$ below\tablenotemark{c}} &
\multicolumn{1}{c}{$K_{yy}$ above\tablenotemark{c}} &
\multicolumn{1}{c}{C$_2$H$_2$\tablenotemark{d}} &
\multicolumn{1}{c}{C$_2$H$_6$\tablenotemark{d}}}
\startdata
 Solar flux & \nodata & \nodata & \nodata &    &                    &                   & 0.34 & 0.34 \\
 Cassini    & \nodata & \nodata & \nodata &    &                    &                   & 0.50 & 0.87 \\
Model A     & Standard& Standard& Standard& 5  & 2$\times$10$^{10}$ & 0                 & 0.49 & 0.77 \\
Model B     & Standard& Standard& Standard& 5  & 2$\times$10$^{10}$ & 2$\times$10$^{9}$ & 0.54 & 0.80 \\
Model C     & Standard& Standard& Standard& 10 & 2$\times$10$^{10}$ & 2$\times$10$^{9}$ & 0.46 & 0.73 \\
Model D     & Standard& Standard& Standard& 10 & 2$\times$10$^{11}$ &                 0 & 0.46 & 0.75 \\
Model E     & Standard& Standard& Standard& 10 & 2$\times$10$^{11}$ & 2$\times$10$^{9}$ & 0.51 & 0.79 \\
Model F     & \citet{Moses et al.2005}& Standard& Standard& 5  & 2$\times$10$^{10}$ & 2$\times$10$^{9}$ & 0.59 & 0.84 \\
Model G     & Standard& $\times$1.1& Standard& 5  & 2$\times$10$^{10}$ & 2$\times$10$^{9}$ & 0.49 & 0.76 \\
Model H     & Standard& Standard& $\times$0.1& 10 & 2$\times$10$^{11}$ & 2$\times$10$^{9}$ & 0.48 & 0.91 \\
\enddata
\tablenotetext{a}{\ Standard chemistry is taken from \citet{Moses et
al.2000}. Standard temperature and $K_{zz}$ profiles are from
\citet{Gladstone et al.1996}. Modified chemistry is from
\citet{Moses et al.2005}. Reference temperature and $K_{zz}$
profiles are set at equator (10$^\circ$). The maximum changes in
temperature (increase by 10\%) and $K_{zz}$ (reduced by a factor of
10) profiles are at 85$^\circ$, with the change assumed to be
linearly proportional to the angle of latitudes.}

\tablenotetext{b}{\ Transition level of $K_{yy}$ profile. The level
is given in units of mbar.}

\tablenotetext{c}{\ Values of $K_{yy}$ below and above the
transition level. $K_{yy}$ is in units of cm$^2$ s$^{-1}$.}

\tablenotetext{d}{\ Ratio of abundance at $\pm$70$^\circ$ latitude
to abundance at the equator. Values for Cassini measurements are
averaged at $\pm$70$^\circ$.}

\end{deluxetable}

\clearpage

\end{document}